\documentclass[aps,prl,reprint,groupedaddress,amsmath,amssymb]{revtex4-2}
\usepackage{graphicx}
\usepackage{xcolor}
\usepackage{hyperref}
\usepackage{tensor}

\definecolor{ao(english)}{rgb}{0.0, 0.5, 0.0}
\definecolor{burntorange}{RGB}{255, 131, 35}

\graphicspath{{./}{Figures/}}

\begin{document}

\title{Collisionless Accretion of Finite-Angular-Momentum Plasma onto a Spinning Black Hole}

\author{John M. Mehlhaff}
\email{mehlhaff@wustl.edu}
\author{Alexander Y. Chen}
\email{cyuran@wustl.edu}
\author{Martin Luepker}
\email{luepker.m@wustl.edu}
\author{Yajie Yuan}
\email{yajiey@wustl.edu}

\affiliation{Physics Department and McDonnell Center for the Space Sciences, Washington University in St.\ Louis; MO, 63130, USA}

\date{\today}

\begin{abstract}
In low-luminosity active galactic nuclei like M87* and Sgr A*, the accretion disk around the central supermassive black hole is tenuous and collisionless. As a result, the usual ideal magnetohydrodynamics (MHD) approximation may not be applicable. In this Letter, we report on the first fully kinetic simulations of the accretion process where the plasma initially has finite angular momentum, focusing on the case of axisymmetry. The simulated accretion flow behaves remarkably similarly to the magnetically arrested disk (MAD) regime of ideal MHD, reproducing episodes of magnetic flux saturation and eruption typical of MADs. The resemblance to fluid models owes largely to kinetic instabilities, which regulate pressure anisotropy in the disk, allowing fluid terms to dominate the angular momentum transfer. In addition, by handling vacuum regions effectively, our kinetic approach probes the matter supply to the jet funnel. We observe no efficient penetration of the accreting material into this region, suggesting that a pair discharge may be required to sustain the Blandford-Znajek process. These initial findings represent important groundwork to be checked and built upon by future kinetic accretion simulations, especially in 3D.
\end{abstract}

\maketitle

\paragraph{Introduction.---}

Black hole (BH) accretion powers some of the most fascinating phenomena in the Universe. Accreting supermassive BHs can form active galactic nuclei (AGN), shining brightly across the electromagnetic spectrum \cite{lynden-bell_1969}. Some AGN even launch powerful, relativistic jets that extend well beyond their host galaxies \cite{blandford_etal_2019}. Understanding the mechanisms underlying BH accretion and how they convert gravitational potential energy into radiation is an important open problem in astrophysics.

Recently, the Event Horizon Telescope (EHT) used interferometry to image the vicinity of some nearby supermassive BHs, allowing us to see the central engine in action. In particular, the EHT images of M87* \cite{eht_m87_i_2019} and Sgr A* \cite{eht_sgra_i_2022} unambiguously identified the BH shadow, and polarization measurements showed strong evidence of ordered magnetic fields near the horizons of both BHs \cite{eht_m87_viii_2021, eht_sgra_viii_2024}. A major scientific enterprise is interpreting the images to constrain properties of the inner accretion flow. This task relies largely on comparison with synthetic images generated from general relativistic magnetohydrodynamics (GRMHD) simulations \cite{eht_m87_viii_2021, eht_sgra_viii_2024, eht_m87_v_2019, eht_sgra_v_2022}.

However, for low-accretion-rate systems such as M87* and Sgr A*, the accreting plasma is collisionless \cite{quataert_etal_2002}. This allows the development of, e.g.,\ nonthermal particle distributions, kinetic plasma instabilities, and multi-temperature components, none of which are captured by GRMHD. For example, to compute radiation from GRMHD simulations, a prescription for the electron temperature inspired by kinetic theory is needed \cite{ressler_etal_2017, chael_etal_2019, yao_etal_2021, chael_2025}. A fully kinetic model would overcome these issues, predicting the accretion dynamics from first principles.

Recently, significant progress has been made toward developing fully kinetic models of BH accretion \cite{galishnikova_etal_2023, vos_etal_2025, figueiredo_etal_2026}. However, these works have so far been limited to the case of Bondi accretion, with zero initial plasma angular momentum. This precludes a description of the more generic case, where orbiting material must shed its angular momentum in order to accrete. In view of these former limitations, the recent identification of a finite-angular-momentum kinetic equilibrium in Kerr spacetime~\cite{luepker_etal_2025} represents a major breakthrough. Such an equilibrium can be used as the starting point for fully kinetic simulations of BH accretion with angular momentum transport. In this Letter, we present the first such simulations.

\paragraph{Methods.---}

We use our GPU-based general relativistic particle-in-cell (GRPIC) code framework \emph{Aperture}~\cite{chen_etal_2025} to perform fully kinetic, 2D axisymmetric simulations. The simulations are conducted in horizon-penetrating spherical Kerr-Schild coordinates,~$x^\mu=(t,r,\theta,\varphi)=(x^0,x^1,x^2,x^3)$ and assume a dimensionless BH spin of~$a=0.998$. Particles are evolved via the general relativistic equations of motion~\cite{bacchini_etal_2018}. Maxwell's equations are solved under a~$3{+}1$ foliation of spacetime~\cite{komissarov_2004} involving: the electric current,~$J^i$, computed from the particles; electric and magnetic fields,~$D^i$ and~$B^i$, measured by local fiducial observers (FIDOs); and auxiliary metric-induced electric and magnetic fields,~$E_i$ and~$H_i$. The numerical grid is logarithmically stretched in~$r$, with~$\Delta (\ln r)$ and~$\Delta \theta$ both constants, and contains~$4096$ cells in each dimension in order to resolve the kinetic scales. More details on our numerical setup and the~$3{+}1$ formalism are given in the End Matter. Throughout this work, Greek (Latin) indices span values~$0$-$3$~($1$-$3$). Unless otherwise stated, all quantities are in Lorentz-Heaviside units.

The simulations start from the Luepker Torus~\cite{luepker_etal_2025}, a stable collisionless plasma equilibrium around a rotating BH. The torus is initialized with respective inner and outer radii~$r_\mathrm{in} = 6.2 r_g$ and~$r_\mathrm{out} \approx 73r_g$ and a fiducial number density peak of~$n_0$ attained near~$r=10 r_g$ on the equatorial plane (see End Matter). Here,~$r_g = GM/c^2$, where~$M$ is the BH mass. Additionally, we initialize a vertical Wald~\cite{wald1974} magnetic field of strength~$B_0$ far from the BH. This field breaks the torus equilibrium by seeding the magnetorotational instability (MRI)~\cite{balbus_hawley_1991, balbus_hawley_1998} and thus kickstarts the accretion process. We tune the torus and magnetic field to achieve a maximum initial~$\beta_{\rm pl} = 2 p / b^2$ value of~$100$ near the point of maximum plasma density. Here,~$p$ and~$b$ are, respectively, the plasma pressure and comoving magnetic field strength. Above and below the equatorial pressure maximum, the vertical extent of the torus spans a few MRI wavelengths,~$\lambda_\mathrm{MRI}$. We simulate a pure $e^\pm$ pair plasma since it eliminates the electron-ion scale separation. In a GRPIC code, there is no need to maintain a numerical floor on the plasma density, so we leave the region outside the torus a complete vacuum.

We perform two contrasting simulations: one with $e^\pm$ pair production, and one without. In the former, we adopt a simplified pair production scheme to mimic photon-photon pair production near the BH: whenever the Lorentz factor $\Gamma$ of an electron or positron in the local FIDO frame reaches a prescribed threshold~$\Gamma_\mathrm{thr}$, it emits a photon of FIDO-measured energy $\epsilon_\mathrm{ph} = 4 m_e c^2$. This photon propagates along a massless geodesic with a fixed probability per timestep of converting into an $e^\pm$ pair. When a photon pair-produces, its energy is evenly split between the two resulting particles. This scheme is identical to what was adopted by~\cite{chen_etal_2025}. We set~$\Gamma_{\rm thr} = 10$: well beyond the typical particle energy in the accretion flow while still lower than energies particles obtain from unscreened horizon-scale electric fields. We also choose the photon mean-free-path~$\ell_\mathrm{ph}\approx 2 r_g$, which yields enough pairs to fuel a spin-powered jet.

\paragraph{Results.---}

In both simulations, the MRI kicks in at a given location after roughly one Keplerian period (its local growth time). This triggers accretion and transforms the torus into a geometrically thick, magnetized, and turbulent accretion disk. Axisymmetric MRI turbulence cannot continue indefinitely and generally decays after~${\sim}10$ local Keplerian orbits \cite{devilliers_hawley_2003, mckinney_gammie_2004, mckinney_2006}. Since our simulations end by~$t=2500r_g/c$ -- about~$12$ orbits at the density maximum -- they probe mostly the pre-decay stage.

\begin{figure}
    \centering
    \includegraphics[width=1.05\linewidth]{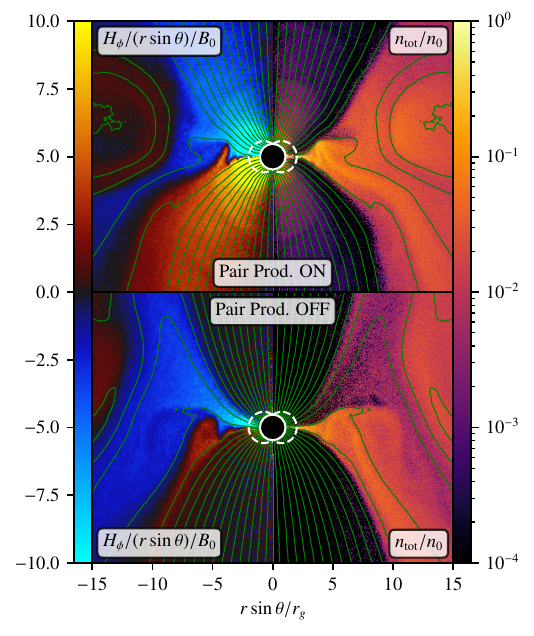}
    \caption{
    Snapshots at~$t=1476 r_g/c$ of the simulations with and without pair production. Panels display the auxiliary magnetic field component $H_\varphi$, and the FIDO-measured total plasma density $n_{\rm tot}$. As shown in the End Matter, in a steady state, $H_\varphi(r,\theta)$ indicates the electric current through the spherical cap defined by~$r$ and~$\theta$.
    }
    \label{fig:global}
\end{figure}

Although weak, the initial vertical magnetic field provides a deep reservoir of magnetic flux. This flux gradually accretes onto the BH and eventually saturates, leading to flux eruptions similar to those of the magnetically arrested disk (MAD) accretion regime~\cite{narayan_etal_2003,igumenshchev_etal_2003,tchekhovskoy_etal_2011,ripperda_etal_2022}. We engineered our setup to reproduce these events by employing a vertical initial field; other initial magnetic configurations may not have concentrated enough flux to saturate and erupt as seen here. Fig.~\ref{fig:global} shows a representative snapshot from each of our simulations, taken at the same time $t = 1476 r_g/c$. Both simulations show a funnel region consisting of quasi-parabolic field lines anchored to the BH as well as a disk that is mostly magnetically disconnected from the BH.

The most striking feature of Fig.~\ref{fig:global} is perhaps the funnel region. In the run without pair production, BH-threading magnetic field lines empty of plasma and particles from the accretion disk do not replenish them (similar to previous findings~\cite{vos_etal_2025} but here seen in the presence of a finite-angular-momentum accretion flow). Such field lines cannot support an azimuthal magnetic field or transmit any Poynting flux. Without pair production, the Blandford-Znajek (BZ) mechanism fails to activate. In contrast, the run with pair production is able to continuously supply the funnel zone with pairs via quasi-periodic discharges~\cite{yuan_etal_2025}. Thus, the BZ mechanism operates, extracting energy from the BH as a Poynting-flux-dominated jet.

\begin{figure}
    \centering
    \includegraphics[width=\linewidth]{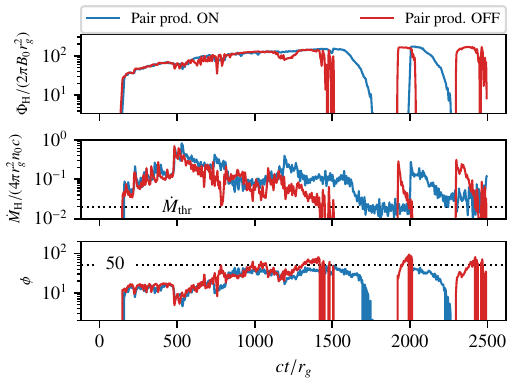}
    \caption{Horizon-penetrating magnetic flux,~$\Phi_{\rm H}$, mass accretion rate,~$\dot{M}_{\rm H}$, and normalized flux,~$\phi = \Phi_{\rm H} (4\pi / (\dot{M}_{\rm H} r_g^2 c))^{1/2}$. Because~$\phi$ becomes unreliable as~$\dot{M}_{\rm H}$ approaches zero, we do not calculate it when~$\dot{M}_{\rm H}$ falls below~$\dot{M}_{\rm thr} = 0.02 (4 \pi r_g^2 n_0 c)$.}
    \label{fig:mdot}
\end{figure}

Fig.~\ref{fig:mdot} shows time series of the magnetic flux, $\Phi_H$, mass accretion rate, $\dot{M}_{\rm H}$, and normalized magnetic flux, $\phi$, through the event horizon for the two simulations. In Lorentz-Heaviside units,~$\phi = \Phi_{\rm H} (4\pi / (\dot{M}_{\rm H} r_g^2c))^{1/2}$; expressions for~$\Phi_{\rm H}$ and~$\dot{M}_{\rm H}$ are given in the End Matter. Both simulations saturate at~$\phi \sim 50$ before a flux eruption is eventually triggered and~$\Phi_{\rm H}$ and~$\dot{M}_{\rm H}$ both plummet -- as in GRMHD simulations of MADs. Such eruptions are also witnessed in kinetic Bondi accretion work~\cite{vos_etal_2025, figueiredo_etal_2026} but are here for the first time fueled by the shedding of angular momentum in a kinetic simulation.

Regardless of whether pair production is active, our simulations show similar saturated~$\phi$-values and hint at similar eruption recurrence times~(${\sim}500 r_g/c$). However, the run with pair production takes significantly longer to expel its horizon magnetic flux due to the presence of plasma in the jet funnel. In this run, magnetic balding occurs through the reconnection of funnel magnetic field lines across an equatorial current sheet, which tends to yield flux decay timescales~${\sim}100 r_g/c$ \cite{crinquand_etal_2021, bransgrove_etal_2021, figueiredo_etal_2026}. Without pair production, even though reconnection is initially triggered in the plasma-filled equatorial current sheet, the upstream funnel field lines contain no particles. As a result, once they have reconnected through the equator, plasma no longer holds the magnetic field in place, allowing the flux to escape the BH within a few~$r_g/c$ \citep{price_1972}.

The most crucial aspect of finite-angular-momentum BH accretion is how the plasma sheds its angular momentum to be able to fall into the BH. We analyze the angular momentum transport in our simulations by isolating contributions from individual stresses following the formalism of \cite{most_etal_2022}. Namely, we split the angular momentum flux term of the stress-energy tensor,~$T^r_\varphi$, as
\begin{equation}
\label{eq:Trphi}
T^r_\varphi = T_{\rm total} = T_\mathrm{plasma} + T_\mathrm{Maxwell} + T_\mathrm{anisotropy} + T_\mathrm{etc},
\end{equation}
where $T_\mathrm{plasma} = (\varepsilon + p + b^2) U^r U_\varphi$ is due to the advection of the plasma, $T_\mathrm{Maxwell} = -b^rb_\varphi$ is the Maxwell stress, $T_\mathrm{anisotropy} = -\Delta p(\hat{b}^r\hat{b}_\varphi - \Delta^r_\varphi/3)$ is the contribution due to pressure anisotropy, and $T_\mathrm{etc}$ accounts for any additional effects. Here,~$\Delta p = p_\perp - p_\parallel$ is the difference between the pressure perpendicular,~$p_\perp$, and parallel,~$p_\parallel$, to the local comoving magnetic field,~$b^\mu$. Full definitions and discussions of these terms are included in the End Matter.

\begin{figure*}
    \centering
    \includegraphics[width=\linewidth]{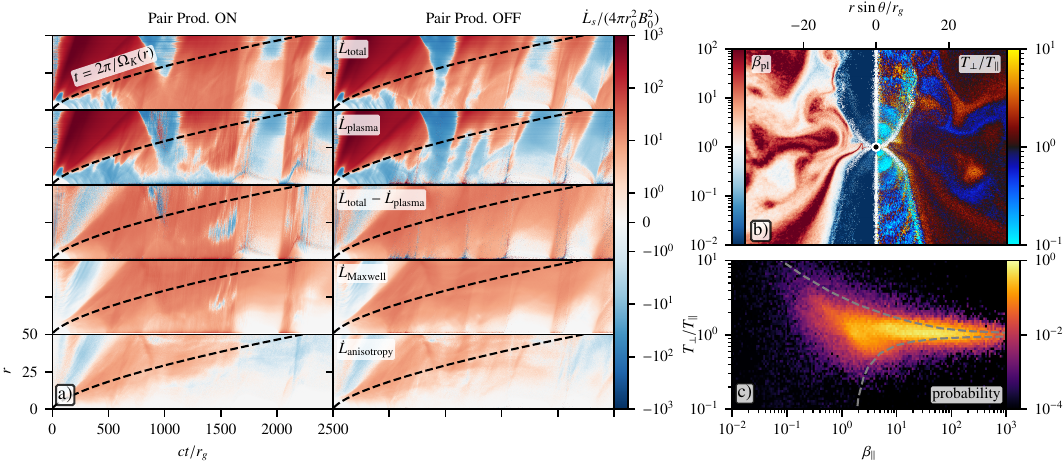}
    \caption{
    Panel a): Spacetime diagrams of the angular momentum fluxes,~$\dot{L}_s$, for the runs with and without pair production. Panel b): Maps, at~$t = 1476r_g/c$, of~$\beta_{\rm pl}$ and the perpendicular-to-parallel temperature ratio,~$T_\perp/T_\parallel = p_\perp / p_\parallel$, for the pair-production simulation. Panel c): Joint probability density of~$\beta_{\parallel}=2p_\parallel/b^2$ and~$T_\perp / T_\parallel$ during the simulation snapshot of panel~b). Following \cite{galishnikova_etal_2023}, dashed lines show where mirror (top) and firehose (bottom) instability growth rates exceed~$10$ per cent of the ion cyclotron frequency. Growth rates are calculated semi-analytically via a method detailed in the supplemental material.
    }
    \label{fig:angmom}
\end{figure*}

In Fig.~\ref{fig:angmom}a, we present spacetime diagrams of the angular momentum flux through constant-$r$ surfaces,~$\dot{L}_{s}(r,t) =2\pi\int_0^{\pi}T_{s}(r,\theta,t) \sqrt{-g}\,d\theta$, where~$s$ denotes the subscripts in Equation~\ref{eq:Trphi}. It can be seen that the plasma contribution,~$\dot{L}_\mathrm{plasma}$, is typically negative, which results from material infall, $U^r < 0$. Negative~$\dot{L}_{\rm plasma}$ first appears at a given radius after roughly one local Keplerian period,~$2\pi/\Omega_K(r)=2\pi (r^3 / G M)^{1/2}$. This is very close to the local MRI growth time, demonstrating the role of the MRI in initiating accretion.

Removing the plasma advection term, the rest of the angular momentum transport,~$\dot{L}_{\rm total} - \dot{L}_{\rm plasma}$, is almost exclusively positive, indicating efficient removal of angular momentum. The Maxwell stress dominates this residual outward angular momentum transport in both simulations. Its magnitude is much higher than the contribution from pressure anisotropy when~$t>2\pi/\Omega_K(r)$.

The left panel of Fig.~\ref{fig:angmom}b shows a spatial map of~$\beta_\mathrm{pl}$ for the pair production run. In the jet funnel,~$\beta_\mathrm{pl}$ is close to $10^{-2}$, but it reaches up to~$10^2$ in the accretion flow. Since~$|T_\mathrm{anisotropy}/T_\mathrm{Maxwell}|$ scales as $|\Delta p|/b^2$ and~$|\Delta p|$ is at most~${\sim}p$, the ratio~$|T_\mathrm{anisotropy}/T_\mathrm{Maxwell}|$ can in principle grow to rival~$p/b^2 \sim \beta_\mathrm{pl}$. Thus, for the Maxwell stress to dominate the angular momentum transfer even in high-$\beta_{\rm pl}$ regions indicates that~$|\Delta p|$ is regulated to~$\ll p$. This can be seen on the right-hand panel of Fig.~\ref{fig:angmom}b, where~$T_\perp/T_\parallel=p_\perp/p_\parallel$ approaches unity in high-$\beta_{\rm pl}$ zones.

The regulation of pressure anisotropy in our simulations is imposed by kinetic mirror and firehose instabilities~\cite{kunz_etal_2016}. These are triggered through the global fluid-level dynamics. For example, differential rotation in the disk amplifies the magnetic field, causing~$T_\perp / T_\parallel$ to grow due to the conservation of particles' magnetic moments, which leads to mirror threshold violations. Once triggered, mirror and firehose modes scatter particle pitch angles, tending to isotropize the distribution function. As a result, the pressure anisotropy remains largely within the thresholds for these instabilities, as shown in Fig.~\ref{fig:angmom}c. Such regulation has been observed in kinetic Bondi accretion models \cite{galishnikova_etal_2023}; this work represents the first kinetic demonstration of its consequences on global angular momentum transport.

Although Figs.~\ref{fig:angmom}b and~\ref{fig:angmom}c show the simulation with pair production, we have checked that the run without pair production exhibits similar~$\Delta p$-regulation. Thus, independently of jet activity, kinetic micro-instabilities limit $T_\mathrm{anisotropy}$ in the accretion disk. Additionally, computational resources restrict the separation between long MHD timescales and short micro-instability timescales in our simulations. We expect the latter to operate even faster compared to global timescales in a real system, all the more firmly enforcing the limits on~$\Delta p$ observed here.

Past 3D works, either involving kinetic shearing-box MRI simulations \cite{bacchini_etal_2024} or global GRMHD simulations with kinetic closures for the pressure anisotropy \cite{foucart_etal_2017, dhruv_etal_2025}, have typically found $T_{\rm anisotropy}\lesssim T_{\rm Maxwell}$, largely consistent with our results. Nevertheless, it will be important to extend our work to 3D. The balance of stresses might change in 3D since the MRI turbulence will likely be far less ordered than in axisymmetry.

\begin{figure}
    \centering
    \includegraphics[trim={0 0 0 22},clip, width=\columnwidth]{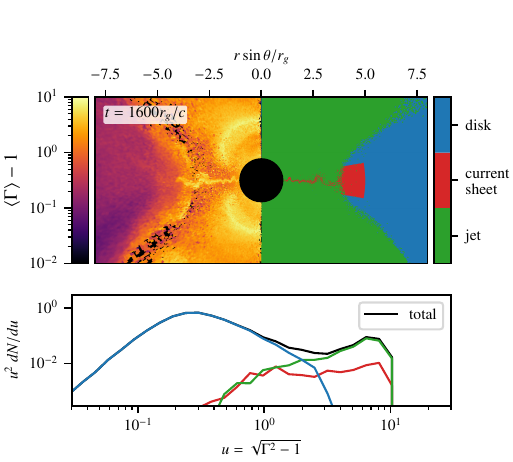}
    \caption{Top: Maps of the local FIDO-measured average Lorentz factor,~$\langle \Gamma \rangle$, and zonal decomposition into disk, current sheet, and jet regions. Bottom: Contributions to the total particle energy distribution from each spatial region. FIDO-measured Lorentz factors~$\Gamma$ and four-velocities~$u$ are related via~$u=\sqrt{\Gamma^2-1}$. The jet is defined by~$\sigma > 0.3$. Where~$\sigma < 0.3$, the current sheet is the wedge~$(r,\theta)\in [r_{\rm H},5r_g] \times [4\pi/9,5\pi/9]$; the disk is everywhere else.}
    \label{fig:ptcldist}
\end{figure}

Finally, we overview the nonthermal particle acceleration in our simulations. Fig.~\ref{fig:ptcldist} shows the total particle energy distribution decomposed into spatial zones -- the jet, the equatorial current sheet, and the accretion disk -- during a flux eruption. Similarly to GRMHD works~\cite{yao_etal_2021}, we define the jet via a magnetization threshold,~$\sigma=b^2/2\rho > 0.3$; we checked that the threshold value can be increased or decreased by a factor of ten without modifying our results. Of the region where~$\sigma<0.3$, we define that in the wedge~$r_{\rm H} < r < 5r_g$ and~$|\theta-\pi/2|<\pi/18$ as the current sheet; the disk is everywhere else. The cutoff at $\Gamma = 10$ in Fig.~\ref{fig:ptcldist} is due to the pair production threshold.

Particle energization in the current sheet is at its maximum during the flux eruption pictured in Fig.~\ref{fig:ptcldist} but still pales compared to that in the jet funnel (where the pair discharge occurs; cf.~\cite{vos_etal_2025}). Importantly, however, particles in the funnel region are accelerated chiefly along magnetic field lines and so attain relatively small pitch angles. Thus, the current sheet, which produces a much broader pitch-angle distribution of energetic particles, could dwarf the high-energy synchrotron output of the funnel despite nominally weaker particle acceleration. By the same token, though, the accelerated current sheet particles are more susceptible to strong synchrotron cooling than those in the funnel. Ultimately, a firmer comparison of the particle distributions and radiative signatures between these two zones requires additional simulations with self-consistent synchrotron radiative cooling.

Compared to the jet and current sheet, we observe a relative dearth of nonthermal particles in the disk zone of our simulations. The distribution remains mostly thermal in the majority of the accretion flow. In contrast, local box simulations of the MRI turbulence have shown that particles can be accelerated to a nonthermal power law~\cite{bacchini_etal_2022, bacchini_etal_2024}. We believe that this tension is due in part to our limited scale separation in these global simulations, as well as to their axisymmetric nature, which
prevents the MRI turbulence from fully developing. Nevertheless, our results highlight the magnetospheric region as the dominant source of the highest-energy particles.

Particle acceleration operates somewhat differently in the absence of pair production. In this case, no jet develops, and the reconnection current sheet becomes charge starved. While this results in a long nonthermal tail in the distribution extending to $\Gamma\gtrsim500$,
the total plasma energization is smaller. 
This is likely due to a lower level of magnetic free energy liberated during the flux eruption event when the polar funnel is vacuum-like.

\paragraph{Conclusion.---}
We have carried out the first global GRPIC simulations of collisionless BH accretion starting from a finite-angular-momentum plasma. In our simulations, the MRI triggers accretion-mediating angular momentum transport -- just like in GRMHD. As the plasma evolves, firehose and mirror instabilities limit its pressure anisotropy, providing an effective collisionality that causes the flow to behave like a magnetized fluid. As a result, fluid stresses, specifically magnetic torque, dominate the angular momentum transport. Our simulations also exhibit magnetic flux saturation and eruption phases typical of the MAD accretion state in GRMHD models.

Besides affirming a GRMHD description of angular momentum transport, our kinetic treatment makes several key advancements beyond fluid frameworks. Our code correctly handles vacuum regions and thus probes the role of pair production in enabling a BZ jet. Absent pair production, our simulations do not exhibit a mechanism for efficiently mixing plasma from the accretion flow into the near-horizon polar regions. Without a plasma to support the necessary current, no BZ jet is launched. This result must be checked in 3D, however, where additional interchange channels might better mix disk plasma into the nascent jet.

A kinetic approach also allows us to model and localize nonthermal particle acceleration. Our simulations show the strongest nonthermal particle acceleration in the jet funnel and equatorial current sheet, with the accretion flow remaining largely thermal. Relaxing the assumption of axisymmetry and enhancing the simulations' dynamic range will enable future work to address key issues left open here: particle acceleration in the accretion disk, mixing of the accreting plasma into the funnel region, and the dominant angular momentum transport channel.

\section{Acknowledgment}
We thank
Fabio Bacchini,
Benoît Cerutti,
Luca Comisso,
Jason Dexter,
Vedant Dhruv,
Alisa Galishnikova,
Matt Kunz,
Sasha Philippov,
Chris Reynolds,
Bart Ripperda,
Navin Sridhar, and
Dimitri Uzdensky
for helpful discussions. JM, ML and YY are supported by a grant from the Simons Foundation (MP-SCMPS-00001470). AC and YY acknowledge support from NSF grants DMS-2235457 and AST-2308111, as well as NASA grant 80NSSC25K0080. AC also acknowledges support from NASA grant 80NSSC24K1095. This work was also facilitated by the Multimessenger Plasma Physics Center (MPPC), NSF grant PHY-2206608 to YY.
This research was supported in part by grant NSF PHY-2309135 to the Kavli Institute for Theoretical Physics (KITP).
This research used resources of the Oak
Ridge Leadership Computing Facility at the Oak Ridge National Laboratory,
which is supported by the Office of Science of the U.S. Department of Energy
under Contract No. DE-AC05-00OR22725. 
The authors also acknowledge the Research Infrastructure Services (RIS) group at Washington University in St. Louis for providing computational resources and services that were used to generate parts of the research results delivered within this paper.

\bibliography{ref.bib}

\clearpage

\section{End Matter}

\subsection{Definitions and Glossary}

This work follows the notation of~\citet{komissarov_2004} for the Kerr metric and 3+1 formulation. We report useful symbol definitions and formulae below. A more complete description of various quantities, including algorithmic implementation details, is given by~\cite{chen_etal_2025}. In this section only, we set~$G=M=c=1$.

The Kerr-Schild-coordinate metric is given by:
\begin{equation}
    \label{eq:kerr-schild-metric}
    \begin{split}
    ds^{2} = &g_{tt}\,dt^{2} + 2g_{t\phi}\,dt d\phi + 2g_{tr}\,dt dr + g_{rr}\,dr^{2}\\
    &+ g_{\theta\theta}\,d\theta^{2} + g_{\phi\phi}\,d\phi^{2} + 2g_{r\phi}\,dr d\phi.
    \end{split}
\end{equation}
Notably, there are cross terms for $t\phi$, $tr$, and $r\phi$. The metric coefficients are:
\begin{equation}
\label{eq:ks-coefficients}
\begin{split}
  g_{tt} &= z - 1,\quad g_{t\phi} = -az \sin^{2}\theta,\quad g_{tr} = z, \\
  g_{rr} &= 1 + z,\quad
  g_{\theta\theta} = \rho^{2},\quad g_{\phi\phi} = \Sigma \sin^{2}\theta / \rho^{2}, \\
  g_{r\phi} &= -a \sin^{2}\theta (1 + z),
\end{split}
\end{equation}
where:
\begin{equation}
    \label{eq:ks-variables}
    \begin{split}
      \rho^{2} &= r^{2} + a^{2}\cos^{2}\theta, \quad z = 2r / \rho^{2},\\
      \Sigma &= (r^{2} + a^{2})^{2} - a^{2}\Delta \sin^{2}\theta,\quad \Delta = r^{2} + a^{2} - 2r. \\
    \end{split}
\end{equation}
The lapse function is~$\alpha = 1 /\sqrt{1 + z}$; the components of the shift vector are~$\beta^r = z/(1 + z)$ and $\beta^\theta = \beta^\phi = 0$. The spatial metric $\gamma_{ij}$ describes the geometry of a spatial slice at constant $t$, and it is simply $\gamma_{ij} = g_{ij}$. However,~$\gamma^{ij} \neq g^{ij}$. The determinants of~$\gamma_{ij}$ and~$g_{\mu \nu}$ are, respectively,~$g$ and~$\gamma$. They are related through~$\sqrt{-g} = \alpha \sqrt{\gamma}$.

The \emph{Aperture} code solves Maxwell's equations in curved spacetime:
\begin{equation}
    \label{eq:D-B-evolution}
    \begin{split}
        \nabla\cdot \mathbf{D} =\rho,&\quad\mathrm{or}\quad \frac{1}{\sqrt{\gamma}}\partial_i \left(\sqrt{\gamma} D^i\right) = \rho, \\
      \nabla\cdot \mathbf{B} =0,&\quad\mathrm{or}\quad \frac{1}{\sqrt{\gamma}}\partial_i \left(\sqrt{\gamma} B^i\right) = 0, \\
     \frac{\partial \mathbf{D}}{\partial t} =\nabla\times\mathbf{H}-\mathbf{J},&\quad\mathrm{or}\quad \partial_t D^i = e^{ijk}\partial_j H_k - J^i, \\
     \frac{\partial \mathbf{B}}{\partial t} =-\nabla\times\mathbf{E},&\quad\mathrm{or}\quad \partial_t B^{i} = -e^{ijk}\partial_{j}E_{k}.
    \end{split}
\end{equation}
The auxiliary fields $E_i$ and $H_i$ are defined as:
\begin{equation}
    \label{eq:E-H-B-D}
    \begin{split}
    \mathbf{E}&=\alpha\mathbf{D}+\pmb{\beta}\times\mathbf{B},\quad\mathrm{or}\quad  E_i = \alpha \gamma_{ij}D^j + e_{ijk}\beta^jB^k, \\
      \mathbf{H}&=\alpha\mathbf{B}-\pmb{\beta}\times\mathbf{D}, \quad\mathrm{or}\quad H_i = \alpha \gamma_{ij}B^j - e_{ijk}\beta^jD^k.
    \end{split}
\end{equation}
Here,
$e_{ijk}$ is the 3-dimensional Levi-Civita pseudo-tensor,
$e_{ijk} = \sqrt{\gamma}\epsilon_{ijk}$ and
$e^{ijk} = \epsilon_{ijk}/\sqrt{\gamma}$, with
$\epsilon_{ijk}$ the Levi-Civita symbol (entries~$1$,~$0$, and~$-1$). The 3-current $J^{i}$ is related to
the 4-current $I^{\mu}$ by $J^{i} = \alpha I^{i}$, and the charge density $\rho$ is $\rho = \alpha I^0$. In the absence of significant displacement current,~$\partial_t \mathbf{D}\simeq0$, the field~$H_\varphi$ indicates the electric current through a spherical cap,
\begin{align}
    2\pi H_\varphi(r,\theta) &= 2\pi \int_0^\theta \partial_{\theta'} H_\varphi(r,\theta') \, \mathrm{d} \theta' \notag \\
    &= 2\pi \int_0^\theta J^r(r,\theta') \sqrt{\gamma}\, \mathrm{d} \theta' \, ,
\end{align}
as mentioned in the caption of Fig.~\ref{fig:global}.

The particles obey the equations of motion~\cite{bacchini_etal_2018}:
\begin{equation}
    \label{eq:particle-eom}
    \begin{split}
      \frac{dx^{i}}{dt} &= \gamma^{ij}\frac{u_{j}}{u^{0}} - \beta^{i}, \\
      \frac{du_{i}}{dt} &= -\alpha u^{0}\partial_{i}\alpha + u_{k}\partial_{i}\beta^{k} - \frac{u_{j}u_{k}}{2u^{0}}\partial_{i}\gamma^{jk} + \frac{F_{i}}{m},
    \end{split}
\end{equation}
where $F_{i}$ is the covariant vector for the Lorentz force: 
\begin{equation}
    \label{eq:Lorentz}
    F_{i} = q\left(\alpha\gamma_{ij}D^{j} + e_{ijk}\gamma^{jl}\frac{u_{l}}{u^{0}}B^{k}\right).
\end{equation}
Photons are treated as massless particles that do not experience the Lorentz force.

Following \cite{luepker_etal_2025}, the distribution function of species~$s$ is
\begin{align}
    f_s(x^i,u_j,t) = \frac{\mathrm{d} N_s}{\mathrm{d} V_x \mathrm{d} V_u} \, ,
\end{align}
where $\mathrm{d} V_x = \sqrt{-g} u^0 \mathrm{d}r \mathrm{d}\theta \mathrm{d}\varphi$, $\mathrm{d} V_u = \mathrm{d} u_r \mathrm{d} u_\theta \mathrm{d} u_\varphi / (u^0 \sqrt{-g})$, and~$g$ denotes the determinant of the metric tensor,~$g_{\mu \nu}$. In our axisymmetric simulations, no quantities depend on~$\varphi$, and so we write~$f(x^i,u_j,t)=f(r,\theta,u_j,t)$.

Given the distribution function, one may define the rest-mass-flux four-vector, 
\begin{align}
    N^\mu(r,\theta, t) = \sum_s m_s\int u^\mu f_s(r,\theta,u_j, t)\, \mathrm{d} V_u \, ,
    \label{eq:nmu}
\end{align}
and stress-energy tensor,
\begin{align}
    T^{\mu\nu}(r,\theta, t) = \sum_s m_s \int u^\mu u^\nu f_s(r,\theta,u_j, t)\, \mathrm{d} V_u \, ,
\end{align}
where~$m_s$ is the rest-mass of a particle of species~$s$.
These are the main objects from which we define below most quantities analyzed in the main text.

The magnetic flux and mass accretion rate through the event horizon are
\begin{align}
    \Phi_{\rm H} &= 2 \pi \int_0^{\pi/2} B^r(r_{\rm H},\theta) \sqrt{\gamma} \, \mathrm{d}\theta \quad \mathrm{and} \\
    \dot{M}_{\rm H} &= -2 \pi \int_0^\pi N^r(r_{\rm H}, \theta) \sqrt{-g} \, \mathrm{d}\theta \, .
\end{align}
Next, following \cite{most_etal_2022}, we define the fluid four-velocity, energy density, pressure, and comoving magnetic field, respectively, as
\begin{align}
    U^\mu &= N^\mu / \sqrt{-N^\nu N_\nu} \, , \notag \\
    \varepsilon &= U_\mu U_\nu T^{\mu \nu} \, , \notag \\
    p &= \frac{1}{3} \Delta_{\mu \nu} T^{\mu \nu} \, , \quad \mathrm{and} \notag \\
    b^\mu &= U_\nu \tensor[^\star]{F}{^\mu^\nu} \, ,
\end{align}
where~$\tensor[^\star]{F}{^\mu^\nu}$ is the dual of the electromagnetic field tensor and~$\Delta^{\mu \nu} = U^{\mu} U^\nu + g^{\mu \nu}$.
In addition, we define parallel and perpendicular pressures,
\begin{align}
    p_\parallel &= \hat{b}_\mu \hat{b}_\nu T^{\mu \nu} \quad \mathrm{and} \notag \\
    p_\perp &= \frac{1}{2} \left( \Delta_{\mu \nu} - \hat{b}_\mu \hat{b}_\nu \right) T^{\mu \nu} \, ,
\end{align}
where~$\hat{b}^\mu = b^\mu / b = b^\mu/(b_\nu b^\nu)^{1/2}$. The total pressure can be written~$p = (2p_\perp + p_\parallel)/3$. The pressure anisotropy is~$\Delta p = p_{\perp} - p_\parallel$. It is bounded by~$-3p \leq \Delta p \leq 3p/2$.

\subsection{Simulation setup}

We initialize the Luepker Torus~\cite{luepker_etal_2025} by specifying four parameters: a fiducial number density scale,~$n_0$; a radial scale,~$r_0$, that corresponds approximately to the location of the number density maximum along the equator; an inner radius,~$r_{\rm in}$; and a so-called ``energy temperature'',~$T$, which dictates how quickly the plasma density decays toward the torus edges. The spatial extent of the torus is determined entirely by~$r_0$ and~$r_{\rm in}$. A broader separation between the two increases the maximum energy of the constituent particles, puffing up the equilibrium vertically and extending it radially. A larger torus stores more plasma fuel for accretion at the cost of increasing the simulation domain size needed to contain it.

We choose~$r_0=12r_g$ and~$r_{\rm in}=6.2r_g$, which places the outer radius at~$r_{\rm out}\approx 73r_g$ and the density maximum at~$r=9.9r_g \sim r_0$. We also select~$T=0.02 m_e c^2$, which roughly matches the difference between the maximum and minimum initial particle energies. Finally, we place the inner and outer simulation boundaries, respectively, at~$r_{\rm min}=r_g$ and~$r_{\rm max}=85r_g$. Choosing~$r_{\rm min}$ below the event horizon at~$r_{\rm H}=1+\sqrt{1-a^2}=1.06r_g$ causally decouples the interior radial boundary from the domain. Our choice of~$r_{\rm max}\gtrsim 1.1r_{\rm out}$ leaves enough space beyond the torus to fit a perfectly matched layer \cite{birdsall_langdon_1991, cerutti_etal_2015}, which absorbs radially outflowing waves and material.

 We then select the remaining parameters --~$B_0$,~$n_0$, and our grid size~$N_r$ -- to satisfy a number of constraints, which we outline below. In what follows, we distinguish the fixed parameters~$B_0$ and~$n_0$ from their spatially dependent counterparts,~$n$ and~$B$ (measured by local FIDOs), by omitting the~$0$ subscript for the latter.
 
 The first constraint we need to satisfy is that the scale height of the torus,~$H$, be at least several multiples of the most-unstable MRI wavelength,~$\lambda_{\rm MRI}=2\pi v_{\rm A}/\Omega_{\rm K}$. Here,~$v_{\rm A}$ is the Alfvén speed, which, for a given magnetization,~$\sigma=B^2/n m_e c^2$, is just~$v_{\rm A}=\sqrt{\sigma/(1+\sigma)} \simeq \sqrt{\sigma}$, with the nonrelativistic expression~$v_{\rm A}\simeq \sqrt{\sigma}$ holding for~$\sigma\ll1$. In practice, we find that it suffices to enforce~$H \gg \lambda_{\rm MRI}$ at~$r_0$; this still creates several MRI wavelengths per scale height across a healthy range of radii. Furthermore, since our torus is geometrically thick, with~$H \sim r$, we phrase this as a requirement on~$r_0$, namely,~$r_0 \gg \lambda_{\rm MRI,0}\equiv 2\pi \sqrt{\sigma_0} / \Omega_{\rm K}(r_0)$. Given that we have already chosen~$r_0=12r_g$, enforcing~$r_0 \gg \lambda_{\rm MRI,0}$ amounts to a choice of~$\sigma_0 = B_0^2 / n_0 m_e c^2$. We select~$\sigma_0=1.8 \times 10^{-4}$, yielding~$\lambda_{\rm MRI,0} / r_0 = 2\pi\sqrt{r_0 \sigma_0 / r_g} = 0.29$.

Next, we must ensure that the MRI operates in a fluid regime, growing on a timescale,~$\tau_{\rm MRI} \simeq 1 / \Omega_{\rm K}(r)$, much slower than the magnetic gyration frequency,~$\omega_B = e B / m_e c$, of individual particles. Satisfying this requirement by a large margin is important in order for kinetic micro-instabilities to operate on much shorter timescales than the fluid-level MRI \cite{bacchini_etal_2022}. Since~$r_0$ is quite close to the most rapidly orbiting inner edge of the disk, and the initial magnetic field strength is approximately uniform and equal to~$B_0$, upholding~$\tau_{\rm MRI} \gg 1/\omega_B$ throughout most of the disk can be achieved by enforcing it only at~$r_0$. We therefore demand that~$\omega_{B_0} \tau_{\rm MRI,0} = e B_0 / (m_e c \Omega_{\rm K}(r_0)) \gg 1$. Introducing the fiducial plasma skin depth,~$d_0 = (m_e c^2 / (n_0 e^2))^{1/2}$, and using the nonrelativistic approximation,~$v_{\rm A} / c = \sqrt{\sigma_0}$, we can rewrite the condition~$\omega_{B_0} \tau_{\rm MRI,0} \gg 1$ as~$\lambda_{\rm MRI,0} / (2 \pi d_0) \gg 1$.

We note that, because~$d_0$ depends only on physical constants and~$n_0$, and since~$\sigma_0 = B_0^2 / n_0 m_e c^2$ has already been selected, the choice of~$\lambda_{\rm MRI,0} / d_0$ fully constrains both~$n_0$ and~$B_0$. Our simulations adopt~$\lambda_{\rm MRI,0} = 160 d_0 \simeq 25 (2\pi d_0)$, corresponding to~$n_0 = 2000 m_e c^2 / (e^2 r_g^2)$ and~$B_0 = 0.6 m_e c^2 / (e r_g)$. Since~$d_0$ is itself resolved by our simulation grid (outlined below), this means that we resolve~$\lambda_{\rm MRI,0}$ with hundreds of grid cells. For comparison, in GRMHD, where it is not necessary to resolve~$d_0$, the most-unstable MRI mode is typically only resolved by tens of grid cells.

Our final constraint, alluded to above, is that the radial grid spacing,~$\Delta r = r\Delta(\ln r) = (r/N_r) \ln(r_{\rm max} / r_{\rm min})$, resolve the plasma skin depth,~$d = (m_e c^2/ (n e^2))^{1/2}$, everywhere. While both quantities vary spatially, we find this condition to be most stringent where the density reaches its maximum of about~$n_0$ near~$r_0$. Thus, we only need to require~$\Delta r|_{r_0} = r_0 \Delta(\ln r) < d_0 = (m_e c^2 / (n_0 e^2))^{1/2}$ in order for the skin depth to be resolved everywhere. This can be translated into the condition
\begin{align}
    N_r &> \frac{r_0}{d_0} \ln \left( \frac{r_{\rm max}}{r_{\rm min}} \right) = \left( \frac{\lambda_{\rm MRI,0}}{d_0} \right) \left( \frac{r_0}{\lambda_{\rm MRI,0}} \right) \ln \left( \frac{r_{\rm max}}{r_{\rm min}} \right) \notag \\
    &= 2500 \left( \frac{\lambda_{\rm MRI,0}}{160 d_0} \right) \left( \frac{r_0}{3.5 \lambda_{\rm MRI,0}} \right) \ln \left( \frac{r_{\rm max}}{80 r_{\rm min}} \right) \, .
\end{align}
To respect this limit, our simulations use~$N_r = N_\theta = 4096$ cells in each dimension.

For an illustrative comparison, we consider a fiducial magnetic field strength for Sgr A* of~$B \sim 30 \, \rm G$ at~$r \simeq 5 r_g$ \cite{eht_sgra_v_2022}. This translates to~$\omega_{B} = 5 \times 10^8 \, \rm s^{-1}$, and, hence,~$\omega_{B} \tau_{\rm MRI} = \omega_{B} / \Omega_K = 1 \times 10^{11}$. Our simulations respect this hierarchy while reducing the dynamic range:~$\omega_B / \Omega_K = \lambda_{\rm MRI} / (2 \pi d_0) = 25$. Future, larger simulations will be able to adopt a broader scale separation, but astrophysical values will surely remain intractable.

\clearpage

\section{Supplemental Material}
\setcounter{page}{1}
\subsection{Simulation software}

We use the GPU-accelerated PIC code \emph{Aperture} designed for studying relativistic plasmas around compact objects~\footnote{https://github.com/fizban007/Aperture4}. The code solves the Maxwell-Vlasov system using the standard Particle-in-Cell algorithm on a spherical grid with logarithmic spacing in the $r$ direction. The electromagnetic field is interpolated to the particle position using a first-order spline, then used to update the particle momenta. Particle trajectories are solved using a ``drift-kick'' approximation, where they are first evolved by a half step according to the geodesic equation, then we apply an electromagnetic push using a modified Boris pusher, and finally they are evolved according to the geodesic equation by another half step. The motion of the particle is used to deposit a current density onto the grid with a first-order shape function using a modified Esirkepov method that conserves charge~\cite{esirkepov_2001,chen_etal_2025}. Finally, the current is used to update the electromagnetic fields using a semi-implicit predictor-corrector iteration scheme. The code has been successfully applied to perform global simulations of the neutron star magnetosphere in multiple works in the past decade~\cite{chen_beloborodov_2014, chen-thesis, chen_beloborodov_2017, chen_etal_2020}. A full description of the code algorithm is contained in reference~\cite{chen_etal_2025}.

\subsection{Pressure anisotropy and the resulting kinetic instabilities}
To obtain the thresholds for mirror and firehose instabilities, we write down the dispersion relation for a plasma with a nonrelativistic bi-Maxwellian distribution at rest in a uniform background magnetic field $\mathbf{B}_0$. We use cgs units in this section. The zeroth order distribution function for species $s$ is
\begin{equation}\label{eq:bi-Maxwellian}
    f_s(v_{\parallel}, v_{\perp})=\left\{\frac{1}{\sqrt{\pi }w_{\|}}\exp \left(-\frac{v_{\|}^2}{w_{\|}^2}\right) \frac{1}{\pi w_{\perp}^2}\exp \left(-\frac{v_{\perp}^2}{w_{\perp}^2}\right)\right\}_s,
\end{equation}
where $v_{\|}$ and $v_{\perp}$ refer to velocities parallel and perpendicular to the background magnetic field, respectively; $w_{\perp s}^2=2k_BT_{\perp s}/m_s$ and $w_{\| s}^2=2k_BT_{\| s}/m_s$, with $T_{\|}$ and $T_{\perp}$ representing the temperatures in the direction parallel and perpendicular to the background magnetic field, respectively, and $k_B$ the Boltzmann constant. The linearized equations for the Fourier components of the electric field perturbations are
\begin{equation}\label{eq:wave_equation}
    \mathbf{k}\times (\mathbf{k}\times\mathbf{E})+\frac{\omega ^2}{c^2}\pmb{\epsilon} \cdot \mathbf{E}=0,
\end{equation}
where the dielectric tensor is
\begin{equation}
    \pmb{\epsilon}(\omega, \mathbf{k})=1+\sum_s \pmb{\chi}_s(\omega,\mathbf{k}),
\end{equation}
and $\pmb{\chi}_s$ is the susceptibility of the s-th plasma component. For the bi-Maxwellian distribution \eqref{eq:bi-Maxwellian}, the susceptibility tensor can be calculated from \cite{stix_1992}
\begin{align}
    \pmb{\chi}_s&=\left(\hat{e}_{\|}\hat{e}_{\|}\frac{2 \omega _{p}^2}{\omega k_{\|}w_{\perp}^2}\left\langle v_{\|}\right\rangle +\frac{ \omega _{\text{p}}^2}{\omega }\sum_{n=-\infty}^{\infty} e^{-\lambda } \mathbf{Y}_n\left(\lambda  \right) \right)_s,
\end{align}
where for each species $s$, the $\lambda$ parameter is
\begin{equation}
    \lambda_s =\frac{k_{\perp}^2w_{\perp s}^2}{2 \Omega_s ^2}=\frac{k_{\perp}^2k_BT_{\perp s}}{m_s \Omega_s ^2},
\end{equation}
$\omega_{p,s}=\sqrt{4\pi n_s q_s^2/m_s}$ is the plasma frequency, and $\Omega_s=q_sB_0/(m_s c)$ is the cyclotron frequency. For distribution~\eqref{eq:bi-Maxwellian}, $\left\langle v_{\|}\right\rangle=0$ for both species, and the tensor $\mathbf{Y}_n(\lambda)$ is
\begin{widetext}
\begin{align}
      \mathbf{Y}_n(\lambda)=\left(
\begin{array}{ccc}
\displaystyle \frac{n^2 I_n}{\lambda}A_n & -i n \left(I_n-I_n'\right)A_n & \displaystyle \frac{k_{\perp}}{\Omega }\frac{n I_n}{\lambda }B_n\\
i n \left(I_n-I_n'\right)A_n & \displaystyle \left(\frac{n^2 I_n}{\lambda }-2 \lambda  I_n'+2 \lambda  I_n\right)A_n & \displaystyle \frac{ik_{\perp}}{\Omega }\left(I_n-I_n'\right)B_n \\
\displaystyle \frac{k_{\perp}}{\Omega }\frac{n I_n}{\lambda }B_n & \displaystyle -\frac{ik_{\perp}}{\Omega }\left(I_n-I_n'\right)B_n & \displaystyle \frac{2 (\omega -n \Omega )}{k_{\|}w_{\perp}^2}I_nB_n
\end{array} 
\right),
\end{align}
\end{widetext}
where $I_n(\lambda)$ is the modified Bessel function of the first kind, and 
\begin{align}
    A_n&=\frac{T_{\perp}-T_{\|}}{\omega T_{\|}}+\frac{1}{k_{\|}w_{\|}}\frac{\left(\omega -n \Omega \right)T_{\perp}+n \Omega T_{\|}}{\omega T_{\|}}Z_0 (\zeta _n),\\
    B_n&=\frac{(\omega -n \Omega) T_{\perp}+n \Omega T_{\|}}{k_{\|}\omega T_{\|}}\left[1+\left(\frac{\omega -n \Omega }{k_{\|}w_{\|}}\right)Z_0(\zeta _n)\right].
\end{align}
Here $\zeta_n=(\omega-n\Omega)/(k_{\|}w_{\|})$, and
\begin{equation}
    Z_0(\zeta)=i\sqrt{\pi}e^{-\zeta^2}(1+\mathrm{erf}(i\zeta))
\end{equation}
is the plasma dispersion function, where $\mathrm{erf}(z)=(2/\sqrt{\pi})\int_0^ze^{-t^2}\,dt$ is the error function.
We write $\mathbf{n}=c\mathbf{k}/\omega$ and, without loss of generality, assume that the background field $\mathbf{B}_0$ is along $z$ and that~$\mathbf{k}$ is in the $x-z$ plane with an angle $\theta$ with respect to $\mathbf{B}_0$. The wave equation~\eqref{eq:wave_equation} then becomes
\begin{equation}\label{eq:wave_equation_component}
     \left(
\begin{array}{ccc}
 \epsilon_{xx}-n_z^2 & \epsilon_{xy} & \epsilon_{xz}+n_x n_z \\
 \epsilon_{yx} & \epsilon_{yy}-n_x^2-n_z^2 & \epsilon_{yz} \\
 \epsilon_{zx}+n_x n_z & \epsilon_{zy} & \epsilon_{zz}-n_x^2 \\
\end{array}
\right)
\left(
\begin{array}{c}
 E_x \\
 E_y \\
 E_z \\
\end{array}
\right)=0.
\end{equation}
The dispersion relation is obtained by requiring the determinant of the coefficient matrix in \eqref{eq:wave_equation_component} to be zero:
\begin{equation}\label{eq:dispersion}
    \mathrm{det}\left(
\begin{array}{ccc}
 \epsilon_{xx}-n_z^2 & \epsilon_{xy} & \epsilon_{xz}+n_x n_z \\
 \epsilon_{yx} & \epsilon_{yy}-n_x^2-n_z^2 & \epsilon_{yz} \\
 \epsilon_{zx}+n_x n_z & \epsilon_{zy} & \epsilon_{zz}-n_x^2 \\
\end{array}
\right)=0.
\end{equation}

For our applications, we consider a pair plasma, so the ion-to-electron mass ratio is $m_i/m_e=1$. We also assume that both species have the same density and temperatures, so $\beta_{\|i}=\beta_{\|e}$ and $T_{\perp i}/T_{\|i}=T_{\perp e}/T_{\|e}$. The dispersion relation is then fully determined by the following parameters: $\beta_{\|i}$, $T_{\perp i}/T_{\|i}$ and the ion magnetization $\sigma_i=B_0^2/(4\pi n_im_ic^2)$. We consider the low-$\sigma_i$ regime as appropriate for the accretion disk. It turns out that as long as $\sigma_i\ll1$, or, more precisely, $k_BT_{\|i}/(m_ic^2)=\beta_{\|i}\sigma_i/2\ll1$, the growth rate of mirror and firehose instabilities does not depend on the exact value of $\sigma_i$. In our calculations, we typically fix $\sigma_i$ at a small value between $10^{-4}$ and $10^{-2}$.

In practice, it appears that the oblique firehose and mirror instabilities constrain the plasma anisotropy (e.g., \cite{hellinger_etal_2006,bale_etal_2009}). These are non-propagating instabilities, meaning that $\omega$ is purely imaginary. To find the growth rate of these instabilities at a given $\beta_{\|i}$, $T_{\perp i}/T_{\|i}$ and $\mathbf{k}$, we employ the \verb|FindRoot| function in \verb|Mathematica| to solve for $\omega$ in the dispersion relation \eqref{eq:dispersion}. We use Newton's method with a purely imaginary starting trial solution, which usually works well as long as the trial solution is not too far from the true solution. Then, for a given set of $\beta_{\|i}$ and $T_{\perp i}/T_{\|i}$, we scan $\mathbf{k}$ in both its magnitude $k$ and its angle $\theta$ with respect to $\mathbf{B}_0$ to obtain the maximum growth rate of the instability. Finally, we find the maximum growth rate as a function of $T_{\perp i}/T_{\|i}$ for a given $\beta_{\|i}$, and locate the $T_{\perp i}/T_{\|i}$ value that gives a maximum growth rate of 10\% the ion cyclotron frequency, which we illustrate in panel c) of Fig.~\ref{fig:angmom} in the main text.

\end{document}